# A dynamical model based on finite stacking enthalpies for homogeneous and inhomogeneous DNA thermal denaturation

Marc JOYEUX[a] and Sahin BUYUKDAGLI

*Laboratoire de Spectrométrie Physique (CNRS UMR 5588),*

*Université Joseph Fourier, BP 87, 38402 St Martin d'Hères, France*

**Abstract** : We present a nonlinear dynamical model for DNA thermal denaturation, which is based on the finite stacking enthalpies used in thermodynamical nearest-neighbour calculations. Within this model, the finiteness of stacking enthalpies is shown to be responsible for the sharpness of calculated melting curves. Transfer-integral and molecular dynamics calculations are performed to demonstrate that the proposed model leads to good agreement with known experimental results for both homogeneous and inhomogeneous DNA.

[a] email : Marc.JOYEUX@ujf-grenoble.fr

Study of the separation of the two strands of DNA upon heating, a phenomenon called *denaturation* or *melting*, is a subject which has a long history [1,2], because it is a preliminary step for the understanding of transcription. From the experimental point of view, the examination of UV absorption spectra of diluted DNA solutions reveals that denaturation occurs through a series of steps as a function of temperature [3]. These jumps suggest that large portions of the inhomogeneous DNA sequence separate over very short temperature intervals, which is of course reminiscent of first-order phase transitions. Various models have been developed along the years to investigate this point. So far, two different mechanisms have emerged, which could potentially be responsible for a first-order phase transition. Peyrard and Bishop (PB) have shown, by studying dynamical models based on non-linear Hamiltonians [4,5], that it is sufficient to consider a weaker stiffness for single stranded DNA compared to double stranded DNA to observe a first-order phase transition [6]. On the other hand, studies relying on the statistical Poland-Scheraga (PS) model [7] lead to the conclusion that excluded volume effects can induce a discontinuous transition, even when stiffness is neglected [8].

The main purpose of this article is to show that the sharpness of the DNA melting transition may also be due to the finiteness of the interaction between stacked base plateaux. This interaction, which originates partly from the overlap of the $\pi$ electrons of the bases and partly from hydrophobic interactions, necessarily vanishes when two successive bases have slid on each other far enough for their rings not to be superposed any longer. This point is correctly taken into account in both the statistical PS models [7] and the nearest-neighbour thermodynamic ones [9], while in the PB models the stacking energy goes to infinity when the separation between the rings goes to infinity. It will precisely be shown, that a Hamiltonian model based on the finite stacking energies, which are routinely and successfully used in thermodynamic calculations [10], displays a sharp first-order-like transition. A further advantage of using thermodynamic stacking energies is that the proposed microscopic model describes homogeneous and inhomogeneous DNA sequences on the same footing and with equal success, as will be shown by comparing experimental and calculated melting curves.

In the PB models [4,5], which have been used to unravel many aspects of melting [6,11,12], DNA is described as a one-dimensional chain, where hydrogen bonds in base pairs (bp) are assumed to be Morse potentials, while stacking energies are taken into account in the form of nearest neighbour potentials :

$$H = \frac{1}{2} m \sum_n \dot{y}_n^2 + D \sum_n \left(1 - e^{-a y_n}\right)^2 + \sum_n W(y_n, y_{n+1}), \tag{1}$$

where $y_n$ represents the transverse stretching of the hydrogen bond connecting the $n^{th}$ pair of bases. PB proposed the potential

$$W(y_n, y_{n+1}) = \frac{K}{2}(y_n - y_{n+1})^2 \left\{1 + \rho e^{-\alpha(y_n + y_{n+1})}\right\}, \tag{2}$$

to describe stacking interactions [5]. They showed that the anharmonic model ($\rho > 0$) leads to denaturation curves in much better agreement with experiment than the harmonic one ($\rho = 0$). It also displays qualitatively different properties [5,6,11], including a first-order phase transition [6]. The tentative explanation is that the variable backbone stiffness introduced by the additional term gives more cooperativity to the melting process. However, the interaction potential of Eq. (2) does not take into account the fact that stacking interactions are necessarily finite. A simple potential, which does so, is $W(y_n, y_{n+1}) = \min(\frac{1}{2}\Delta H, \frac{1}{2}K(y_n - y_{n+1})^2)$, where $\frac{1}{2}\Delta H$ is the (finite) stacking energy. This is the first interaction potential we used and all the results presented below also hold for this potential (in particular, it leads to very sharp denaturation curves and to singularities in the temperature evolution of the specific heat and the entropy, which validates the conclusion drawn below that the finiteness of stacking interactions may be responsible for the sharpness of melting curves). However, we later modified it for two reasons. First, we replaced the min( ) function by a more physical exponential switching function. Moreover, for this potential, the thermal separation of the two strands is accompanied by the breaking of each strand into many small pieces, that is, one observes large $y_n - y_{n+1}$ values in addition to large $y_n$ ones. This artifact was cancelled by adding a small term, which models the stiffness of the sugar/phosphate backbones. The interaction potential we propose is thus of the form

$$W(y_n, y_{n+1}) = \frac{\Delta H}{2}\left(1 - e^{-b(y_n - y_{n+1})^2}\right) + K_b(y_n - y_{n+1})^2 \ . \tag{3}$$

It is emphasized that the numerical value which will be assumed for $K_b$ is very small (2000 times smaller than PB's $K$ parameter), so that the corresponding term keeps $y_n - y_{n+1}$ to reasonable values without affecting the melting dynamics. Note also that $y_n$ is no longer considered as a pure stretching coordinate in Eq. (3). Indeed, unstacking of two consecutive bases certainly involves some rotation of the bases in the planes perpendicular to the strands. Therefore, $y_n$ is meant to describe in some global manner the separation of two bases located on opposite strands.

In this work, numerical values of the parameters for the PB models are those of Refs. [4,5], that is, $m$=300 amu, $D$=0.04 eV, $a$=4.45 Å$^{-1}$, $K$=0.04 eV Å$^{-2}$, $\rho$=0.5 and $\alpha$=0.35 Å$^{-1}$, while we have assumed $\Delta H$ =0.44 eV, $b$=0.10 Å$^{-2}$ and $K_b$=10$^{-5}$ eV Å$^{-2}$ in Eq. (3) ($b$ has been chosen, such that the quadratic term in the expansion of Eq. (3) is very close to PB's one, that is, $\frac{1}{2}K$). Fig. 1 displays the evolution of $W(y_n, y_{n+1})$ as a function of $y_n - y_{n+1}$ for these numerical values. It is seen that the finiteness of the stacking interaction leads to a well-marked plateau, in spite of the additional quadratic term.

The melting curve for the model of Eq. (3) is compared in Fig. 2 to the curves for the harmonic and anharmonic PB models. The scattered symbols show the temperature evolution of the average bond length $\langle y \rangle$ obtained with the transfer-integral (TI) method in the thermodynamic limit of infinitely long chains (see Appendix A for calculation detail). It is seen that the melting of the proposed model is as sharp as the melting of the anharmonic PB model. Since there is some debate on the applicability of the TI method to models with bound on-site potentials [11], molecular dynamics (MD) simulations were also performed to check the results of TI calculations (see Appendix A for calculation detail). The melting curve for the proposed model obtained from MD simulations applied to a 2399 bp sequence with periodic boundary conditions is displayed in Fig. 2 as a solid line. It fully confirms the results obtained with the TI method. The model of Eq. (3) therefore shows that the finiteness of stacking interactions may be responsible for the sharpness of melting curves.

Most importantly, the value $\frac{1}{2}\Delta H = 0.22$ eV is of the same order of magnitude as the stacking, or "propagation", enthalpies, which are routinely used in thermodynamics calculations. For example, Table 1 of Ref. [10] shows that the amount of energy, which is released upon extending an existing helix by one additional base pair, varies between 0.347 eV and 0.465 eV, depending on the involved nucleotides (note that these quantities are for two base pairs). This indicates that the model of Eq. (3) is quite sensible and can straightforwardly be extended to describe inhomogeneous DNA, according to

$$
\begin{aligned}
H = & \frac{1}{2} m \sum_n \left( \dot{u}_n^2 + \dot{v}_n^2 \right) + D \sum_n \left( 1 - e^{-a(u_n - v_n)} \right)^2 \\
& + \frac{1}{2} \sum_n \Delta H^{(n)} \left( 2 - e^{-b(u_n - u_{n+1})^2} - e^{-b(v_n - v_{n+1})^2} \right) \\
& + K_b \sum_n (u_n - u_{n+1})^2 + K_b \sum_n (v_n - v_{n+1})^2
\end{aligned}
\quad (4)
$$

In this equation, $u_n$ and $v_n$ represent the displacements from equilibrium of bases located on opposite strands, while $\Delta H^{(n)}$ is the amount of energy which is released when extending the existing helix (or ladder, in this case) from the $n^{th}$ to the $(n+1)^{th}$ base pair. The model of Eq.

(3), where $\Delta H$ is eventually replaced by the $\Delta H^{(n)}$, is obtained upon approximate separation of Eq. (4) into the problem for the hydrogen bond stretches $y_n = (u_n - v_n)/\sqrt{2}$ and the centres of mass displacements $x_n = (u_n + v_n)/\sqrt{2}$ [4]. The ten $\Delta H^{(n)}$ values, which are necessary to model a particular DNA sequence, are borrowed from Table 1 of Ref. [10], while the other parameters are taken to be $m$=300 amu, $D$=0.05 eV, $a$=3.5 Å$^{-1}$, $b$=0.05 Å$^{-2}$ and $K_b$=10$^{-5}$ eV Å$^{-2}$. Fig. 3 shows the results of MD simulations performed for three different 2399 bp DNA sequences with open ends, namely, a ATATAT... sequence, a GCGCGC... sequence, and an inhomogeneous sequence found in the litterature [13]. This figure displays the temperature evolution of the probability for the distance $u_n - v_n$ to be larger than 15 Å (a threshold value of 7 Å leads to very similar plots). It is seen that the model of Eq. (4) successfully reproduces three principal properties of experimental denaturation curves, that is, (i) narrow transition temperature ranges, (ii) a gap slightly larger than 40 K between the melting temperatures of pure AT sequences and pure GC ones, and (iii) melting temperatures proportional to the GC percentage. Indeed, the sequence found in the litterature contains 61% of GC bonds, which correlates well with the position of its melting curve.

Moreover, the model of Eq. (4) also reproduces the multi-step process, which is experimentally observed for DNA sequences of length ≈1000-10000 bp. This is most easily checked in Fig. 4. The top plot of this figure displays, for increasing temperatures, the average value $\langle y_n \rangle$ as a function of the site number $n$ for a 1793 bp inhomogeneous sequence found in the litterature (NCBI entry code NM_001101). The profile at each temperature is averaged over hundred 10 ns MD simulations. The bottom plot of Fig. 4 shows, as a function of $n$, the AT percentage averaged over 40 consecutive base pairs of the same sequence. This percentage globally increases from $n$=1 to $n$=1793, with a minimum around $n$=150 and three local maxima around $n$=1300, $n$=1450 and $n$=1600. Examination of the top plot shows that melting correspondingly starts in narrow regions centred around these three later values, while the sequence abruptly melts for all $n \geq 1100$ at slightly higher temperatures. There is then a plateau of about 4 K before the lower end of the sequence abruptly melts around 330 K. Not surprisingly, the portion of the sequence which melts at the highest temperature is located around the AT percentage minimum at $n=150$. These melting domains are in excellent agreement with those obtained from statistical mechanics calculations (see Fig. 2 of Ref. [14]).

Figs. 3 and 4 unambiguously show that, within the model of Eq. (4), the variation of stacking energies is sufficient to explain the variation of melting temperatures for different DNA sequences, while preceding studies based on the PB models relied instead on different parameters for the hydrogen bond Morse potentials and neglected the differences in stacking energies [11,15]. Unfortunately, the Morse potential parameters appear to be more poorly determined than the stacking energies - in part because *one* Morse potential is expected to model the *two or three* different hydrogen bonds which connect two bases -, so that they vary widely from one study to the other [11,15]. Nonetheless, it will certainly be interesting to take the difference in Morse potentials into account in Eq. (4) when the values of the parameters are more firmly established.

Let us now turn back to the model of Eq. (3) and its implications concerning the dynamics of DNA melting. Fig. 5 shows the temperature evolution of the entropy per site $s$ and the specific heat $c_V$ of homogeneous DNA according to the PB models and the model of Eq. (3), as obtained from the TI method. The first-order (or very narrow second-order) phase transition of the anharmonic PB model is revealed by the divergence of $c_V$ and the discontinuity of $s$ at the melting temperature. Although the curves look slightly different, the model of Eq. (3) displays similar features, as can be checked in Fig. 5. The larger the range of $y$ values used in TI calculations, the narrower the switching temperature interval in the plot of $s$ and the higher the peak in the plot of $c_V$ (we checked that the maximum of $c_V$ increases linearly with the maximum $y$ value). Although more involved calculations would be necessary to ascertain this point [6], this strongly suggests that the melting of DNA is also associated with a first-order phase transition for the model of Eq. (3). MD simulations were again performed to check the TI results. Indeed, $c_V$ can be estimated from MD computations as the derivative of the average internal energy of the sequence with respect to temperature, that is, in $k_B$ units, $c_V = \dfrac{1}{Nk_B}\dfrac{d\langle H\rangle}{dT}$ for a sequence of length $N$. The solid line in Fig. 6 shows the temperature evolution of the specific heat obtained from MD simulations, while the circles show the results of TI calculations. Despite the poorer temperature resolution of MD simulations compared to TI calculations, the two curves are in good agreement.

At that point, it should be realized that, although the expressions for $W(y_n, y_{n+1})$ are different, the physical reason why the PB model and that of Eq. (3) lead to a first order (or very narrow second-order) phase transition is the same for both models. It results indeed from a strong entropic effect due to the significant softening of the DNA chain at large displacements. Whether this softening is due, in real systems, to a weaker stiffness of single stranded DNA compared to double stranded DNA, as in the anharmonic PB model, to the finiteness of stacking interactions, as in the models of Eqs. (3) and (4), or to a combination of these two possibilities, remains an open question.

MD calculations were also used to investigate the influence of the length of the sequence on the melting dynamics (TI calculations are of no help for this purpose because this requires the knowledge of an unattainable number of eigenvalues of the TI operator). The temperature evolution of $c_V$ for a 100 bp sequence with open ends is shown as a dashed line in Fig. 6. This curve is essentially displaced by about 20 K to lower temperature compared to the 2399 bp sequence. Although the magnitude of the shift is too large, this displacement qualitatively agrees with experiment. The most important point, however, is that Fig. 6 suggests that there is no fundamental difference between the denaturation dynamics of relatively short and rather long sequences.

To summarize, we have shown that the finiteness of stacking interactions may be responsible for the sharpness of DNA denaturation curves. The proposed microscopic model, which explicitly uses the stacking enthalpies of thermodynamical calculations, leads to good agreement with known experimental results for both homogeneous and inhomogeneous DNA, and suggests, like the anharmonic PB model, that DNA melting is associated with a first-order phase transition. In future work, it will be interesting to investigate whether this kind of model can be adapted to describe thinner details of the rich DNA dynamics.

**APPENDIX A : Transfer-Integral and Molecular Dynamics calculations.**

The Transfer-Integral (TI) method consists in finding the eigenvalues $\lambda_k$ and eigenvectors $\phi_k$ of the symmetric TI operator, which satisfy

$$\int \phi_k(x) \exp\left(-\frac{1}{k_B T}\left\{\frac{1}{2}D\left(1-e^{-ax}\right)^2 + \frac{1}{2}D\left(1-e^{-ay}\right)^2 + W(x,y)\right\}\right) dx = \lambda_k \phi_k(y). \qquad (5)$$

For a sequence of length $N$, the mean stretching of hydrogen bonds is then obtained as $\langle y \rangle = \sum_k (\lambda_k^N \int y \phi_k^2(y) dy) / \sum_k \lambda_k^N$, the entropy per site (in $k_B$ units) as $s = \frac{1}{2} + \frac{1}{2}\ln(2\pi m k_B T) + \frac{1}{N}\frac{\partial}{\partial T}(T \ln \sum_k \lambda_k^N)$ and the specific heat at constant volume (in $k_B$ units) as $c_V = \frac{1}{2} + \frac{1}{N}T\frac{\partial^2}{\partial T^2}(T \ln \sum_k \lambda_k^N)$. In the thermodynamic limit of infinitely long chains ($N \to \infty$), the sums reduce to the term associated with the largest eigenvalue $\lambda_0$ and the corresponding eigenvector $\phi_0(y)$. The $\phi_k(y)$ and $\lambda_k$ were obtained numerically, according to the matrix diagonalization procedure described in Appendix B of Ref. [16]. The diagonalized matrices were of size 2800*2800, corresponding to 2800 $ay$ values regularly spaced between $ay = -10$ and $ay = 3000$.

On the other hand, Molecular Dynamics (MD) simulations consist in integrating numerically Langevin's equations

$$m\frac{d^2 a_n}{dt^2} = -\frac{\partial H}{\partial a_n} - m\gamma \frac{da_n}{dt} + (2m\gamma k_B T)^{1/2} w(t) \ , \tag{6}$$

where $a_n$ stands for $y_n$, $u_n$ or $v_n$, $\gamma$ is the dissipation coefficient (assumed value is $\gamma$=5 ns$^{-1}$), and $w(t)$ a normally distributed random function with zero mean and unit variance. The second and third term in the right-hand side of Eq. (6) model the effect of the solvent on the DNA sequence. Step by step integration (10 fs steps) was performed by applying a second order BBK integrator [17] to the system initially at equilibrium at 0 K and subjected to a temperature ramp of 20 ns/K. This "slow" ramp insures that the average temperature of the system, calculated from its average kinetic energy, closely follows the temperature of the random kicks. Various statistical quantities, like the average energy $\langle H \rangle$, the average bond length $\langle y \rangle$, and the probability for the bond lengths to be larger than given thresholds, were averaged over 0.5 K temperature intervals, that is, 10 ns time intervals.

**FIGURE CAPTIONS**

<u>Figure 1</u> : (Color online) Plot of the stacking energy $W(y_n, y_{n+1})$ (in eV) as a function of $y_n - y_{n+1}$ (in Å).

<u>Figure 2</u> : (Color online) Variation of the mean bond length $\langle y \rangle$ vs temperature. Scattered symbols show the results of TI calculations in the thermodynamic limit of infinitely long chains for the harmonic PB model [4] (crosses), the anharmonic PB model [5] (squares), and the model of Eq. (3) (circles). The solid line shows the result of MD simulations for the model of Eq. (3) and a 2399 bp sequence with periodic boundary conditions.

<u>Figure 3</u> : (Color online) Denaturation curves, averaged over 4 MD runs, for the model of Eq. (4) and three 2399 bp sequences with open ends : an ATATAT... sequence (left), an inhomogeneous sequence with 61% GC base pairs [13] (middle), and a GCGCGC... sequence (right). The vertical scale indicates the probability $p(u_n - v_n > 15 \text{ Å})$ in percents.

<u>Figure 4</u> : (Color online) (Top) Plot, for increasing temperatures, of $\langle y_n \rangle$ as a function of the site number *n* for a 1793 bp inhomogeneous sequence found in the litterature (NCBI entry code NM_001101). (Bottom) Plot, as a function of *n*, of the AT percentage averaged over 40 consecutive base pairs of the same sequence.

<u>Figure 5</u> : (Color online) Variation of the entropy per site *s* and the specific heat $c_V$ (both in $k_B$ units) vs temperature, obtained from TI calculations in the thermodynamic limit of infinitely long chains. Shown are the results for the harmonic PB model [4] (crosses), the anharmonic PB model [5] (squares), and the model of Eq. (3) (circles).

Figure 6 : (Color online) Variation of the specific heat $c_V$ (in $k_B$ units) *vs* temperature for the model of Eq. (3). Circles show the result of TI calculations in the thermodynamic limit of infinitely long chains, the solid line the result of MD simulations applied to a 2399 bp sequence with periodic boundary conditions (average over 4 runs), and the dashed line the result of MD simulations applied to a 100 bp sequence with open ends (average over 80 runs).

**Figure 1**

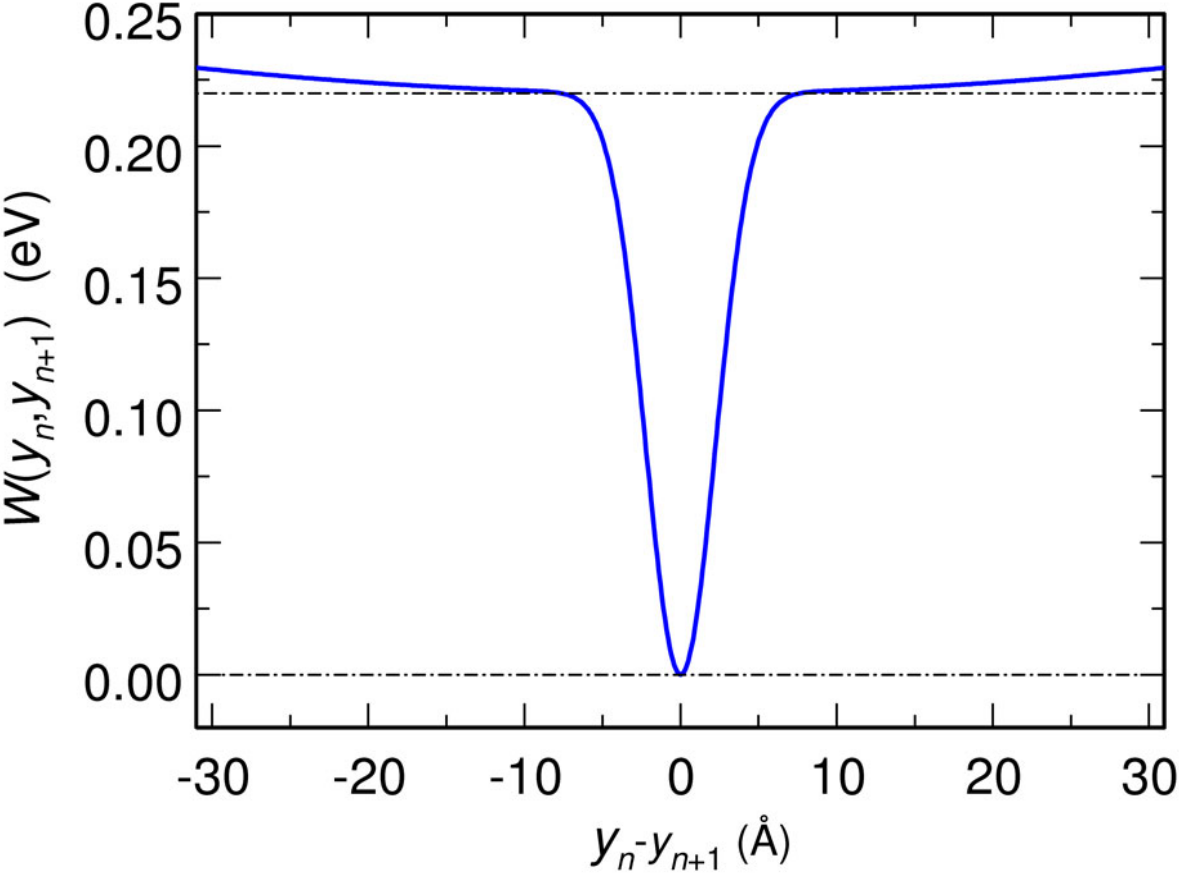

**Figure 2**

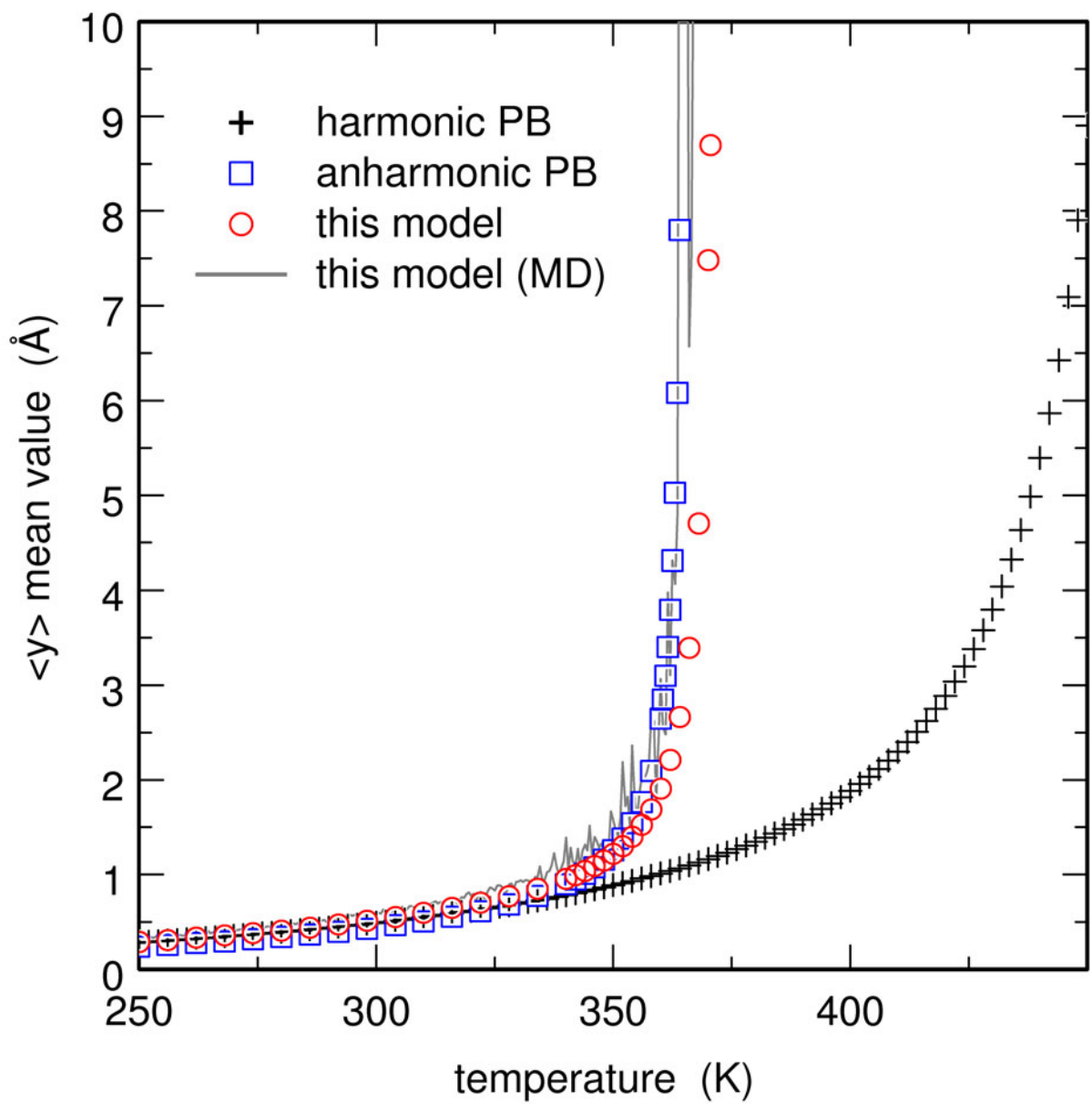

**Figure 3**

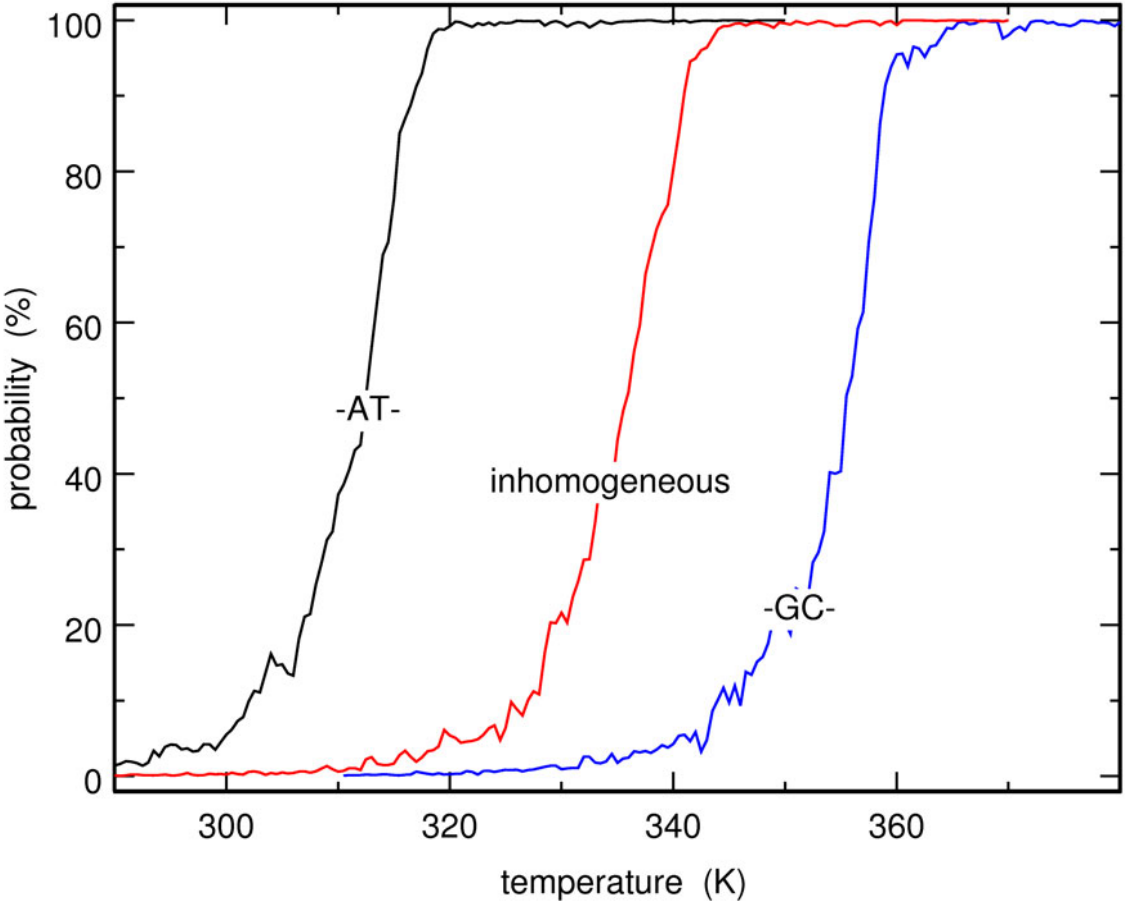

**Figure 4**

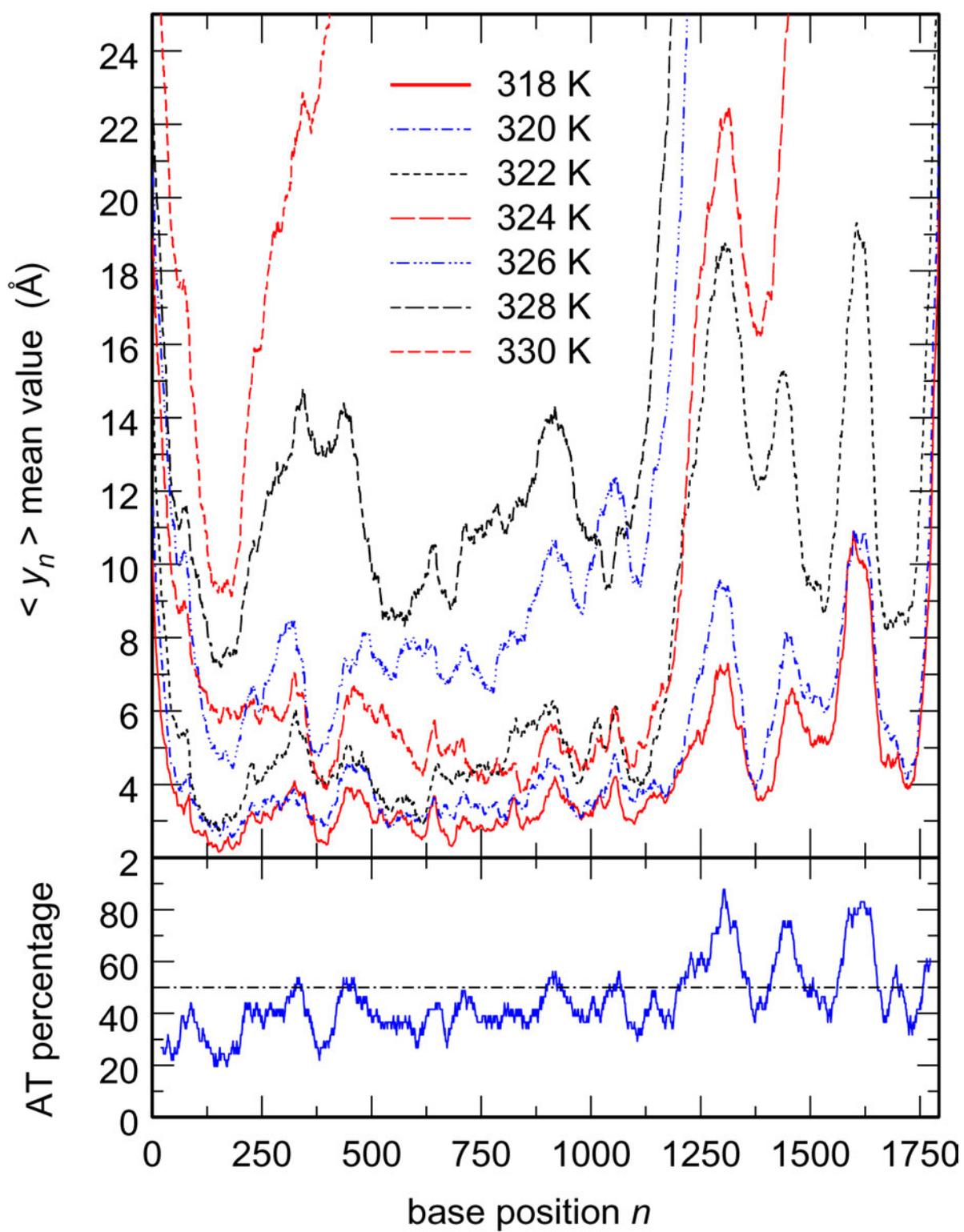

**Figure 5**

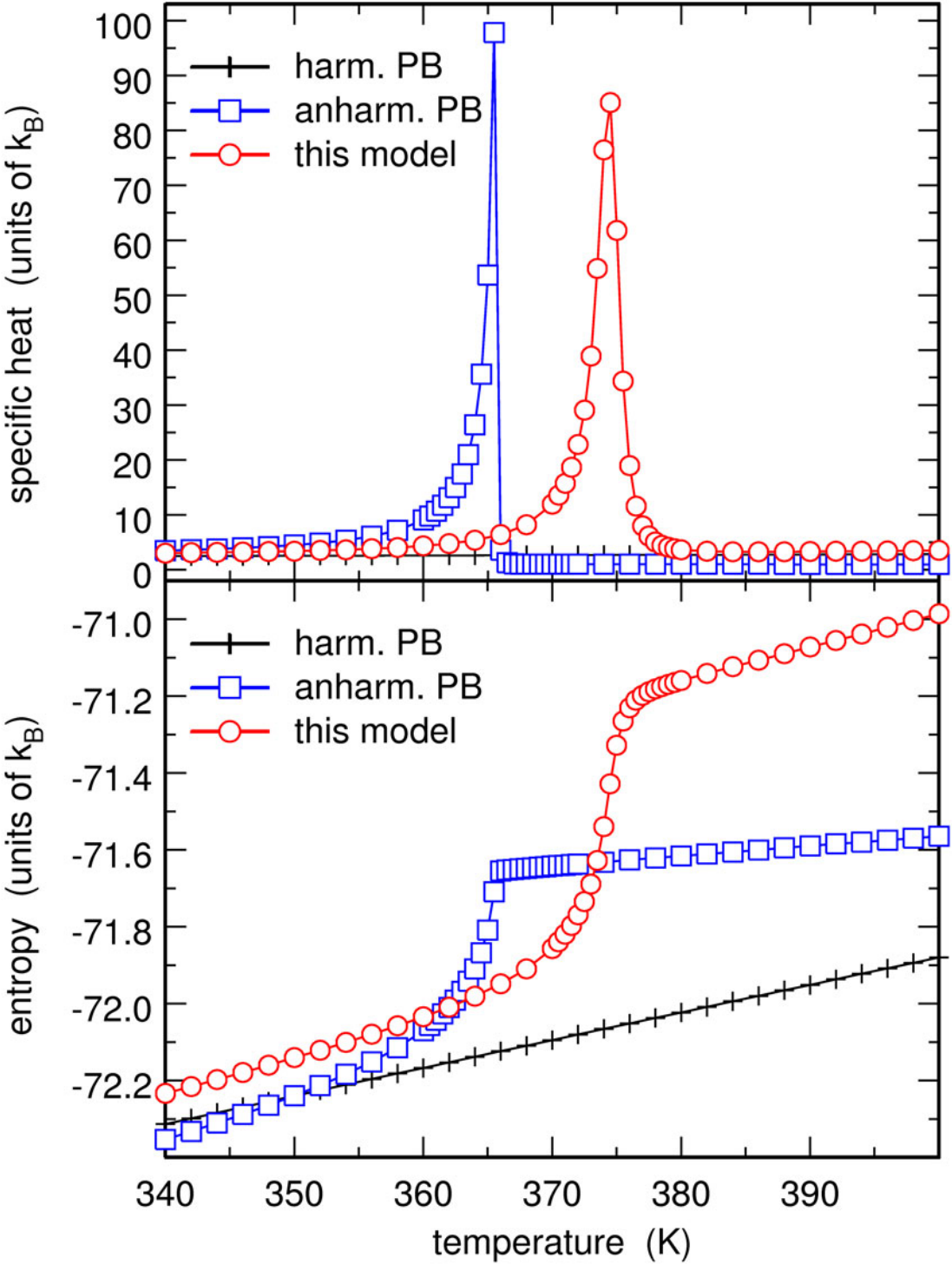

**Figure 6**

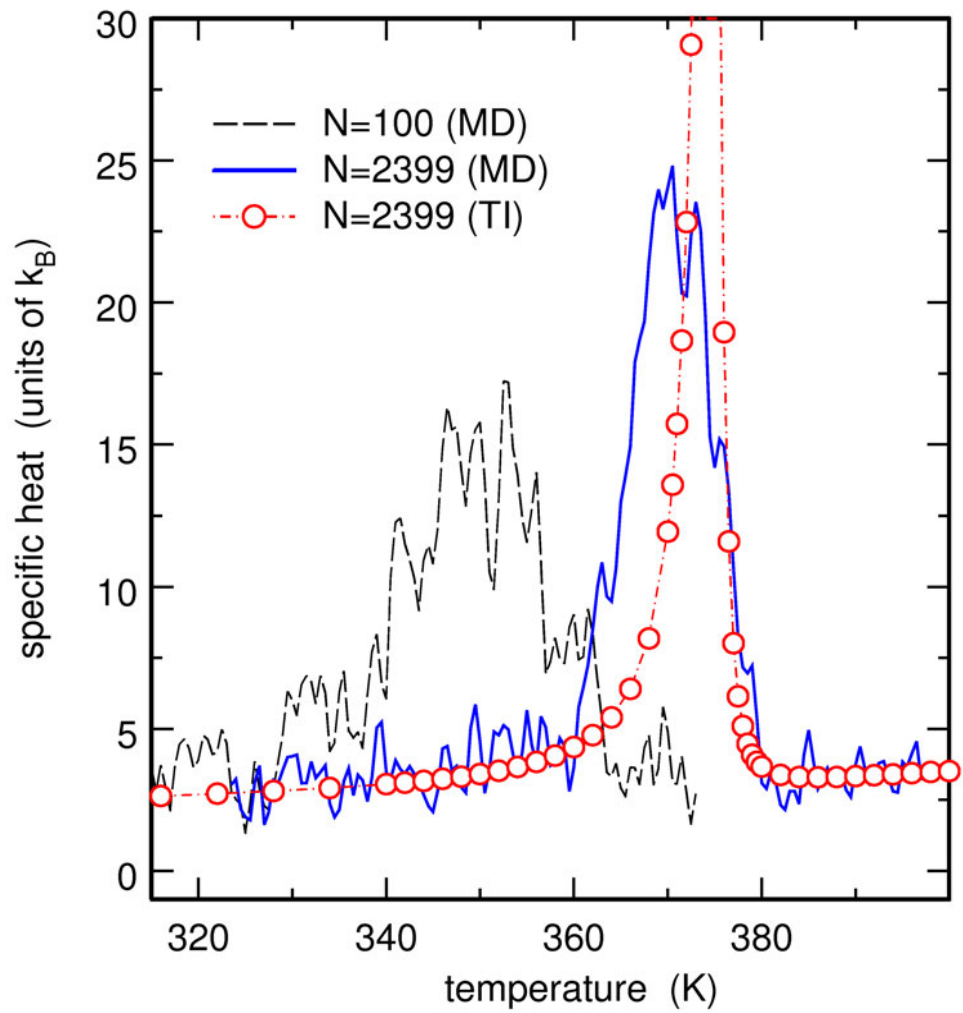